\documentstyle{article}
\begin{document}
%




\def\lsk{\lineskip10pt}

%
%
\catcode`\@=11
%
%
%
\def\@citex[#1]#2{%
\if@filesw \immediate \write \@auxout {\string \citation {#2}}\fi
\@tempcntb\m@ne \let\@h@ld\relax \def\@citea{}%
\@cite{%
  \@for \@citeb:=#2\do {%
    \@ifundefined {b@\@citeb}%
      {\@h@ld\@citea\@tempcntb\m@ne{\bf ?}%
      \@warning {Citation `\@citeb ' on page \thepage \space undefined}}%
      {\@tempcnta\@tempcntb \advance\@tempcnta\@ne%
      \@tempcntb\number\csname b@\@citeb \endcsname \relax%
      \ifnum\@tempcnta=\@tempcntb 
        \ifx\@h@ld\relax%
          \edef \@h@ld{\@citea\csname b@\@citeb\endcsname}%
        \else%
          \edef\@h@ld{\ifmmode{-}\else--\fi\csname b@\@citeb\endcsname}%
        \fi%
      \else
        \@h@ld\@citea\csname b@\@citeb \endcsname%
        \let\@h@ld\relax%
      \fi}%
    \def\@citea{,\penalty\@highpenalty\,}%
  }\@h@ld
}{#1}}

%
\def\@citeb#1#2{{[#1]\if@tempswa , #2\fi}}
%
%
\def\@citeu#1#2{{$^{#1}$\if@tempswa , #2\fi }}
%
%
\def\@citep#1#2{{#1\if@tempswa , #2\fi}}

%
%
\def\bcites{         
        \catcode`\@=11
        \let\@cite=\@citeb
        \catcode`\@=12
}

\def\upcites{         
        \catcode`\@=11
        \let\@cite=\@citeu
        \catcode`\@=12
}

\def\plaincites{      
        \catcode`\@=11
        \let\@cite=\@citep
        \catcode`\@=12
}

%
%

\makeatletter
\@addtoreset{equation}{section}
\renewcommand{\theequation}{\thesection.\arabic{equation}}
\makeatother

\begin{center}  
{\Large {\bf Casimir Effect in {\LARGE $\kappa$}-deformed Theory

}}
\vspace*{15mm}
\vspace*{1mm}

{\bf Soonkeon Nam${^1}$, Heeyong Park and Yunseok Seo$^2$}

\vspace*{0.4cm}
{\it {Department of Physics and Research Institute for Basic Sciences,\\
Kyung Hee University, Seoul, 130-701, Korea}}

\vspace*{0.4cm}

{\tt ${}^1$nam@khu.ac.kr,$\,$${}^2$seo@string.khu.ac.kr}

\vspace*{1cm}
\end{center}



\begin{abstract}
Quantum mechanical fluctuations in an interval give rise to the Casimir effect, which destabilizes the size of the interval. This can be problematic in constructing Kaluza-Klein theories. We consider the possibility that a breakdown of the Poincar\'e symmetry in an extra dimension can solve this instability problem. As a specific example, we consider the space-time with a $\kappa$-deformed Poincar\'e algebra, calculate the Casimir force between two plates, and find that we can have an interval with a stable size.
\end{abstract}


\section{Introduction}
The existence of a large extra dimension may provide new insights into problems such as the hierarchy problem \cite{dimo,dimo2,dimo3,randal,LY}. In that scenario \cite{randal}, only gravity can propagate in the higher dimensional bulk, and all other fields are confined to a lower dimensional brane. The background solution in Ref.\cite{randal} consists of two parallel flat branes - positive tension and negative tension- embedded in a five-dimensional Anti-de Sitter(AdS) bulk. Recently, other scenarios in which additional fields propagate in the bulk have been introduced \cite{gher,davou,bagger}.\par
When the fields propagate in the bulk, a Casimir force which appears from vacuum fluctuation  between two plates, causes a quantum instability of system \cite{casimir}. Therefore, the Casimir force makes the size of the extra dimension go to zero or to infinity, thus finding the stability of a two plate system becomes important. \par
There are some papers which find the stabilization of two branes. Goldberger and Wise \cite{wise} proposed that a radion effective potential may stabilize two branes. Garriga at al. \cite{garriga,garriga2} calculated the radion effective potential $V_{eff} (\phi)$ due to the Kaluza-Klein graviton and other massless bulk fields and found that the radion field $\phi$ could be stabilized through the Casimir force induced by the bulk field. The phenomenological study of the one-loop effect of the radion is in \cite{KKP}. However, there are some problems in providing a theoretical basis for the origin of the radion effective potential. \par
Brevik et al.\cite{brevik} calculated the thermodynamic energy (effective potential) at non-zero temperature. They suggested a new dynamical mechanism to stabilize the thermal brane-world universe in the low energy limit. In that paper, a more natural hierarchy was generated at low temperature compared with that of the hot inflationary universe.\par
Another candidate to stabilize the size of the extra dimension was suggested in Ref.\cite{nam}, where the non-commutativity of space time in the extra dimension was considered. In noncommutavive theory, the commutation relations among space-time coordinates are given by
\begin{equation}\label{comm1}
\lbrack x^{\mu}, x^{\nu} \rbrack = i \theta^{\mu \nu},
\end{equation}
and from the uncertainty relations,
\begin{equation}
\Delta x^{\mu} \Delta x^{\nu} \ge {1 \over 2} |\theta^{\mu \nu} |,
\end{equation}
we know that $\theta^{\mu \nu}$ introduces a minimum area in the $\mu$, $\nu$ plane. The space-time which satisfies Eq.(\ref{comm1}) is called the canonical noncommutative space-time. Experimental tests for the noncommutativity of $\theta^{\mu \nu}$ from deformed dispersion relations of canonical noncommutative space-time have been proposed in Ref.\cite{ANS}. From this noncommutativity, Gomis et al.\cite{gomis} calculated a Kaluza-Klein spectrum in a compact noncommutative extra dimension. The Casimir effect from a one-loop quantum correction in a compact noncommutative extra dimension was obtained in Ref.\cite{nam}, where the next order correction due to the noncommutativity was found to give either an attractive or a repulsive Casimir force. Also a stabilization of the size of the extra dimension in a gravity field was proposed. \par
In this paper, we explore another type of deformed Poincar\'e symmetry. The lesson to be learned in the approach is that a deformation of the Poincar\'e symmetry can be a source of stability for a two brane world. One example is $\kappa$-deformed theory. Using the relativistic harmonic oscillator Hamiltonian in $\kappa$-deformed space-time, we will calculate the Casimir force for a two parallel plate system. 

\section{Casimir effect}
To set the notation, we briefly review the Casimir effect in ordinary space-time. Consider a massless scalar field $\phi$ confined between two parallel plates separated by a distance $a$ \cite{casimir,nam,milton}. Using the Dirichlet boundary condition

\begin{equation}
\phi (0) =\phi (a)=0,
\end{equation}
the Casimir energy per unit area can be obtained by summing up the zero-point energies:

\begin{equation}
u={1 \over 2} \sum_\alpha \omega_\alpha = {1 \over 2} \sum_{n=1}^{\infty} \int {{d^d k \over {(2 \pi)^d}} \sqrt{k^2 + {n^2 \pi^2 \over a^2}}}.
\end{equation}
We set $\hbar=c=1$. The integers $n=1,2,\cdots$ label the normal modes between the plates, and $k$ is the transverse momenta along the plate. The sum is formally divergent.  \par
If the definition of the Gamma function\cite{nam,milton},
\begin{equation}
\Gamma (z) = p^z \int_0^{\infty} dt \, e^{pt} t^{z-1},
\end{equation}
is used, the energy per unit area becomes
\begin{eqnarray}
u=&&{1 \over 2} \sum_{n=1}^{\infty} \int {d^d k \over {(2 \pi)}^d} {1 \over {\Gamma (-{1\over 2})}}\cr
  && \times \int_0^{\infty} {dt \over t} t^{-1/2} e^{-t(k^2 + n^2 \pi^2 /a^2)}.
\end{eqnarray}
After performing a Gaussian integration and an integration with respect to $t$, and using the definition of the Riemann zeta function, we can obtain the energy per unit area:
\begin{equation}
u=-{1 \over {4 \sqrt \pi}}{1 \over (4 \pi)^{d/2}} \left({\pi \over a}\right)^{d+1} \Gamma \left(-{d+1 \over 2}\right) \zeta (-d-1).
\end{equation}
To avoid infinity when $z$ is negative even integer, we use the reflection formula as follows:
\begin{equation}
\Gamma\left({z \over 2}\right) \pi^{-z/2} \zeta{z} = \Gamma \left({1-z \over 2}\right) \pi^{(z-1)/2} \zeta (1-z).
\end{equation}
The the final result is
\begin{equation}
u=-{1 \over {a^{d+1}}} \Gamma \left({d+2 \over 2}\right) (4 \pi)^{-(d+2)/2} \zeta (d+2).
\end{equation}

The energy is always negative in even dimensions. In the $d=2$ case, the Casimir energy per unit area in the transverse direction is
\begin{equation}
u=-{\pi^2 \over 1440} {1 \over a^3}.
\end{equation}
Then, we can obtain the force per unit area between the plates by taking the negative derivative of the energy per unit area with respect to $a$,
\begin{equation}
f=-{\partial u \over {\partial a}} = -{3 \over 1440} {\pi^2 \over a^4}.
\end{equation}
In even dimensions $d$, the Casimir force is always attractive; therefore, the two plate system cannot be stable. This system spontaneously collapses to one plate.

\section{Casimir effect in {\Large $\kappa-$}deformed theory}
Since the $\kappa$-deformed Poincar\'e algebra was introduced in Ref.\cite{lukierski} and \cite{lukierski2}, it has been widely studied in many papers theoretically\cite{nowicki,lukierski3,majid,lukierski4}, and phenomenologically \cite{amelino,amelino2}. In $\kappa$-deformed Minkowski space, which is called a Lie-algebra noncommutative space-time, the commutation relations are given by\cite{amelino2}
\begin{eqnarray}
\lbrack \hat x_0, \hat x_i \rbrack &=& i \lambda \hat x_i, \cr
\lbrack \hat x_i, \hat x_j \rbrack &=& 0,
\label{commutation}
\end{eqnarray}
where $\lambda=1/\kappa$. Since we set $\hbar = c = 1$, $\lambda$ has the dimension of length. \par
From the commutation relations in Eq.(\ref{commutation}), we obtain a $\kappa$-deformed Poincar\'e algebra, written in the bi-cross-product basis \cite{majid,lukierski4} which consists of a classical Lorentz algebra
\begin{eqnarray}
\lbrack M_{\mu \nu}, M_{\rho \tau} \rbrack =\!\!\!\!\! &&i \,(g_{\mu \rho} M_{\nu \tau} + g_{\nu \tau} M_{\mu \rho}- g_{\mu \tau} M_{\nu \rho} \cr
&&- g_{\nu \tau} M_{\mu \tau} ), 
\end{eqnarray}
and a deformed commutation relations
\begin{eqnarray}
\lbrack M_i , P_j \rbrack &=& i\, \epsilon_{ijk} P_k, \cr
\lbrack M_i , P_0 \rbrack &=& 0, \cr
\lbrack N_i , P_j \rbrack &=& -i \delta_{ij} \left( {1 \over {2 \lambda}} (1-e^{2 \lambda P_0}) + {\lambda \over 2}(\vec P)^2 \right) + i \lambda P_i P_j, \cr
\lbrack N_i , P_0 \rbrack &=& i P_i ,
\end{eqnarray}
where $M_i = {1 \over 2} \epsilon_{ijk} M_{jk}, N_i = M_{i0}$. \\
In a $\kappa$-deformed Poincar\'e algebra, the mass Casimir takes the form\cite{amelino2}
\begin{equation}
C_2^{\lambda} = {\vec P}^2 e^{- \lambda P_0} -\left({2 \over \lambda} \sinh {\lambda P_0 \over 2} \right)^2 = -M^2.
\end{equation}
In the case of a massless scalar field\,($M=0$) localized between two plates, the dispersion relation can be written as 
\begin{equation}
P_0={1\over\lambda}\ln{\left(1+\lambda \sqrt{{\vec P}\,^2}\,\, \right)} .
\end{equation}
Since $P_0 = i{\partial \over \partial t}$\cite{lukierski4}, we can write down the Schr\"odinger equation for a massless scalar field $\Psi$ and find a relation between the energy in $\kappa$-deformed space-time and the energy in ordinary space-time:
\begin{equation}
P_0\Psi\,=\,i{\partial\over\partial t}\Psi = H\Psi\,=\,E^\lambda\Psi,
\end{equation}
where $E^\lambda$ is the $\kappa$-deformed energy. By putting $\sqrt{ {\vec P}^2}$ into the non-deformed Hamiltonian $H_0$ and substituting the dispersion relation into the Schr\"odinger equation, we can relate the $\kappa$-deformed energy to the non-deformed energy $E_0$ as follows:
\begin{eqnarray} 
{1\over\lambda} \ln(1+\lambda\sqrt{{\vec P}\,^2}\,\,)\,\Psi&=&{1\over\lambda} \ln(1+\lambda H_0)\,\Psi \cr
E^\lambda&=&{1\over\lambda} \ln(1+\lambda E_0).
\end{eqnarray}
To calculate the Casimir effect, we choose a non-deformed Hamiltonian for harmonic oscillator. Then, the harmonic oscillator energy of in $\kappa$-deformed space-time is
\begin{equation}
E^\lambda_n = {1\over\lambda}\ln \left[ 1+\lambda\omega\left(n+{1\over 2}\right) \right].
\end{equation}

It is easy to show for the ordinary space-time that in the limit  $ \lambda\rightarrow 0$, the $\kappa$-deformed energy goes to the ordinary harmonic oscillator energy:
\begin{equation}
\lim_{ \lambda\rightarrow 0}E^\lambda_n=\omega\left(n+{1\over 2}\right).
\end{equation}
Because the Casimir effect comes from the zero-mode energy, we expand the zero-mode energy (for very small $\lambda$, $\lambda \ll 1$):
\begin{eqnarray}
E^{\lambda}_0 &=& {1\over 2}\omega-{1\over 8}\lambda\ \omega^2+{1\over 24}\lambda^2 \omega^3-\ldots \cr
&=& u_1 + u_2 + u_3 +\cdots ,
\end{eqnarray}
where $\omega=\sqrt{{\vec k}\,^2}$. We divide $\omega$ into two parts - one perpendicular to the  plates and one along the plates. Since the direction perpendicular to the plates is compactified, the momentum is quantized as follow:
\begin{eqnarray}
{\vec k}\,^2 &=& {\vec k_\parallel}^2+{\vec k_\perp}^2 \cr
&=& {\vec k_\parallel}^2 + \Big({n \pi \over a}\Big)^2.
\end{eqnarray}
Then, we can calculate the Casimir energy by extracting the finite terms. \par
The first term is the same as the ordinary Casimir energy in the previous section,
\begin{eqnarray}
u_1&=&{1\over 2}\sum^\infty_{n=-\infty}\int{d^dk\over (2\pi)^d}\sqrt{{\vec k}\,^2+\left({n\pi\over a}\right)^2}\cr
&=&-{1\over a^{d+1}}(4\pi)^{-{(d+2)\over 2}}\Gamma\left({d+2\over 2}\right)\zeta (d+2).
\end{eqnarray}
The second term becomes
\begin{eqnarray}
u_2&=&-{\lambda\over 8}\sum^\infty_{n=-\infty}\int{d^dk\over (2\pi)^d}\,\left[\,\vec k^2+\Big({n\pi\over a}\Big)^2\,\right] \cr
&=&-{\lambda\over 8}\Biggl[\int{d^dk\over (2\pi)^d}\,\vec k^2\,+\,\int {d^d k \over {(2 \pi)^2}} {\pi\over a^2}\zeta(-2)\Biggr].
\label{u2}
\end{eqnarray}
In Eq.(\ref{u2}), only the second part depends on $a$. We can set this part to zero because the zeta function vanishes. We will apply this result to higher even powers of $\omega$. The third term is similar to the first term, but has a different sign
\begin{eqnarray}
u_3&=&{\lambda^2\over 24}\sum^\infty_{n=-\infty}\int{d^dk\over (2\pi)^d}\,\left[\,\vec k^2+\Big({n\pi\over a}\Big)^2\,\right]^{3 \over 2} \cr
&=&{\lambda^2\over 128}(4\pi)^{-{(d+2)\over 2}}{1\over a^{d+3}}\Gamma\left({d+4\over 2}\right)\zeta(d+4).
\end{eqnarray}
Finally, the total Casimir energy is 
\begin{eqnarray}
u_{tot}=&-&({4\pi})^{-{(d+2)\over 2}}{1\over a^{d+1}}\Gamma\left({d+2\over 2}\right)\zeta(d+2)\cr
&-&{\lambda\over 8}\int{d^dk\over (2\pi)^d}\,\vec k^2 \cr
&+&{\lambda^2\over 128}(4\pi)^{-{(d+2)\over 2}}{1\over a^{d+3}}\Gamma\left({d+4\over 2}\right)\zeta(d+4) \cr
&+&{\cal O}(\lambda^3).
\label{total}
\end{eqnarray}

The force per unit area between the plates can be obtained by taking the negative derivative of energy respect to $a$:
\begin{eqnarray}\label{force}
f&=&-{\partial u \over \partial a} \cr
&=&-(4\pi)^{-{(d+2)\over 2}}\left({d+1\over a^{d+2}}\right)\Gamma\left({d+2\over 2}\right)\zeta(d+2)\cr
&&+{\lambda^2\over 128}(4\pi)^{-{(d+2)\over 2}}\left({d+3\over a^{d+4}}\right)\Gamma\left({d+4\over 2}\right)\zeta(d+4) \cr
&&+\,{\cal O}(\lambda^3).
\end{eqnarray}
The first term in Eq.(\ref{force}) is the ordinary Casimir force which causes the quantum instability. The second term in Eq.(\ref{total}) is independent of $a$, so it does not contribute to the Casimir force. The signature of the last term in Eq.(\ref{force}) is always positive in even dimensions. Therefore, it provides a repulsive force. \par
In even dimensions, odd power of $\lambda$ terms give a repulsive force, and the even power of $\lambda$ terms give an attractive force. For small $\lambda$, we can ignore the higher order terms in Eq.(\ref{force}). The different parts of the Casimir force in the leading-order term (attractive and repulsive) can provide stability to a two plate system.\par
From the result of this equation, we can calculate a numerical value of the separation between the two plates. For example, in the case of $d=2$, the ratio between $a$ and $\lambda$, which makes the two-plate system stable, is 
\begin{equation}
a \sim 0.14 \lambda.
\end{equation}
The value of $a$ approaches $\lambda$ as $d$ increases, but it always smaller than $\lambda$. Because the size of noncommutativity of $\lambda$ is very small (about the Plank scale), it is convenient to introduce a very small extra dimension.

\section{Discussion}
Here, we calculated the Casimir force in $\kappa$-deformed theory and found stability for a small extra dimension. There are several papers which suggest stability for an extra dimension \cite{nam,gomis} by using the Kaluza-Klein mode from quantum corrections. However, in this paper, the Casimir effect we derived was from the commutation relation, not from quantum correction. Thus, we can say that it is possible to stabilize an extra dimension on a classical level. \par
However, some problems exist in $\kappa$-deformed theory. The commutation relation in $\kappa$-deformed theory is different from the commutation relation of ordinary noncommutative space-time so called a canonical commutation relation. $\kappa$-deformed theory has a Lie-algebra-like commutation relation. The physical meaning of this theory needs further study. Another interesting problem is the study of free-field theory in $\kappa$-deformed space-time \cite{lukierski5}. There are no well-defined $\kappa$-deformed second-quantized field theories yet.


\section*{Acknowledgements}
SN acknowledges grant R01-2000-000-00021-0 from the Korean Science and Engineering Foundation.


\end{document}